\documentclass[prb,twocolumn,amsmath,amssymb,floatfix, superscriptaddress]{revtex4-1}

\usepackage{color}
\usepackage{xcolor}
\usepackage{tabularx}

\usepackage{graphicx}
\usepackage[caption = false]{subfig}
\usepackage{tikz}
\usetikzlibrary {shapes.geometric}
\usetikzlibrary{decorations.markings}

\usepackage{bm}
\usepackage{float}
\usepackage{dsfont}

\usepackage{overpic}

\newcommand{\ket}[1]{\left| #1 \right. \rangle}

\newcommand{\eps}{\varepsilon}

\newcommand{\kp}{$\mathbf{k \cdot p} $ }

\begin{document}
\title{Band structure and end states in InAs/GaSb core-shell-shell nanowires}

\author{Florinda Vi\~nas Bostr\"om}
\affiliation{Division of Solid State Physics and NanoLund, Lund University, Box 118, S-221 00, Lund, Sweden}
\author{Athanasios Tsintzis}
\affiliation{Division of Solid State Physics and NanoLund, Lund University, Box 118, S-221 00, Lund, Sweden}
\author{Michael Hell}
\affiliation{Division of Solid State Physics and NanoLund, Lund University, Box 118, S-221 00, Lund, Sweden}
\author{Martin Leijnse}
\affiliation{Division of Solid State Physics and NanoLund, Lund University, Box 118, S-221 00, Lund, Sweden}

%%%%%%%%%%%%%%%%%%%%%%%%%%%%%%%%%%%%%%%%%%%%%%
%%%%%%%%%%%%%%%%%%%%%%%%%%%%%%%%%%%%%%%%%%%%%%
%%%%%%%%%%%%%%%%%  Abstract %%%%%%%%%%%%%%%%%%%%%%%
%%%%%%%%%%%%%%%%%%%%%%%%%%%%%%%%%%%%%%%%%%%%%%
%%%%%%%%%%%%%%%%%%%%%%%%%%%%%%%%%%%%%%%%%%%%%%

\begin{abstract}
Quantum wells in InAs/GaSb heterostructures can be tuned to a topological regime associated with the quantum spin Hall effect, which arises due to an inverted band gap and hybridized electron and hole states. Here, we investigate electron-hole hybridization and the fate of the quantum spin Hall effect in a quasi one-dimensional geometry, realized in a core-shell-shell nanowire with an insulator core and InAs and GaSb shells. We calculate the band structure for an infinitely long nanowire using \kp theory within the Kane model and the envelope function approximation, then map the result onto a BHZ model which is used to investigate finite-length wires. Clearly, quantum spin Hall edge states cannot appear in the core-shell-shell nanowires which lack one-dimensional edges, but in the inverted band-gap regime we find that the finite-length wires instead host localized states at the wire ends. These end states are not topologically protected, they are four-fold degenerate and split into two Kramers pairs in the presence of potential disorder along the axial direction. However, there is some remnant of the topological protection of the quantum spin Hall edge states in the sense that the end states are fully robust to (time-reversal preserving) angular disorder, as long as the bulk band gap is not closed.
\end{abstract}
\maketitle

%%%%%%%%%%%%%%%%%%%%%%%%%%%%%%%%%%%%%%%%%%%%%%
%%%%%%%%%%%%%%%%%%%%%%%%%%%%%%%%%%%%%%%%%%%%%%
%%%%%%%%%%%%%%%%%  Intro %%%%%%%%%%%%%%%%%%%%%%%
%%%%%%%%%%%%%%%%%%%%%%%%%%%%%%%%%%%%%%%%%%%%%%
%%%%%%%%%%%%%%%%%%%%%%%%%%%%%%%%%%%%%%%%%%%%%%

\section{Introduction}
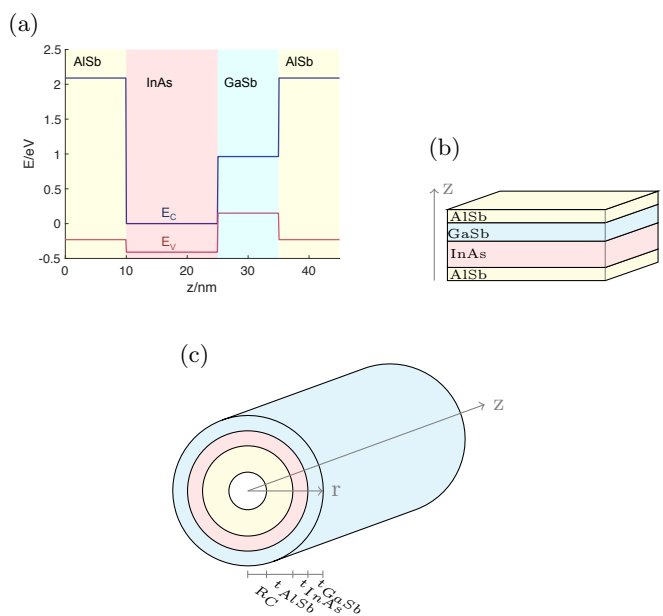
\begin{figure}[t]
\begin{overpic}[width=0.3\textwidth,trim={5cm 3cm 5cm 3cm},clip]{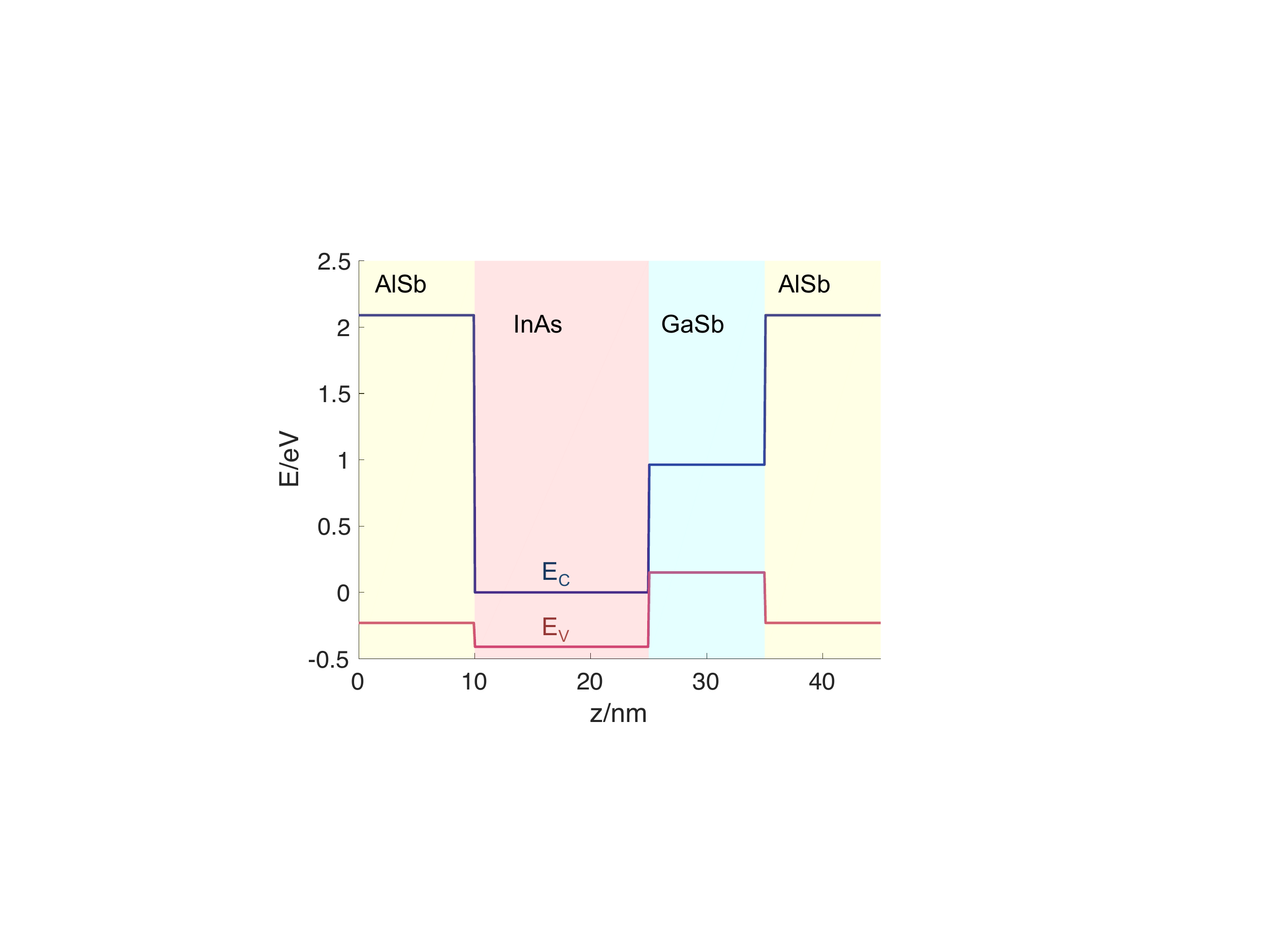}
\put(0,75){(a)}
\end{overpic}
\begin{tikzpicture}[scale=0.35]
\draw[] (-0.1,5)  node [black] {(b)};
\draw[white,-] (0,-2) -- (3,-2) ;

\draw[gray,->] (-0.5,0) -- (-0.5,3.5) node [at end, right, gray] {z};

\filldraw[fill=yellow!13, draw=black] (0,0) rectangle (6,0.5);
\filldraw[fill=pink!40!, draw=black] (0,0.5) rectangle (6,1.5);
\filldraw[fill=cyan!10, draw=black] (0,1.5) rectangle (6,2.2);
\filldraw[fill=yellow!13!, draw=black] (0,2.2) rectangle (6,2.7);
\filldraw[fill=yellow!13!, draw=black] (6,0) -- (8,0.7) -- (8,1.2) -- (6,0.5);
\filldraw[fill=pink!40, draw=black] (6,0.5) -- (8,1.2) -- (8,2.2) -- (6,1.5);
\filldraw[fill=cyan!10, draw=black] (6,1.5) -- (8,2.2) -- (8,2.9) -- (6,2.2);
\filldraw[fill=yellow!13!, draw=black] (6,2.2) -- (8,2.9) -- (8,3.4) -- (6,2.7);
\filldraw[fill=yellow!13!, draw=black] (0,2.7) -- (6,2.7) -- (8,3.4) -- (2,3.4) -- cycle;
\useasboundingbox[black, font=\tiny] (0.8,0) -- (0.8,3) node[pos=0.08]{AlSb} node[pos=0.34]{InAs} node[pos=0.60]{GaSb}  node[pos=0.81]{AlSb};
\end{tikzpicture}\\\
\begin{tikzpicture}[scale=1]
\draw[] (-2.6,1.1)  node [black] {(c)};
\node [cylinder, black, fill=cyan!10, rotate=200, draw, inner sep=1cm, aspect=1.0] (c) {};
\draw[fill=pink!40, font=\small] (-1.89,-0.69) circle (0.8)  node[anchor=south west] {};
\draw[fill=yellow!13, font=\small] (-1.89,-0.69) circle (0.6)  node[anchor=south west] {};
\draw[fill=white, font=\small] (-1.89,-0.69) circle (0.25)  node[anchor=south west] {};

\draw[gray,->] (-1.89,-0.69) -- (2*1.89/3,2*0.69/3) node [at end, right, gray] {z};
\draw[gray,->] (-1.89,-0.69) -- (-0.89,-0.69) node [at end, right, gray] {r};
%\draw[gray,<->] (-1.89,-1.8) -- (-1.89+0.4,-1.8) node [below, gray] {$R_C$};
\draw[gray] (-1.89,-1.8) -- (-0.89,-1.8) node [below, gray] {};
\draw[gray] (-1.89,-1.8+0.05) -- (-1.89,-1.8-0.05) node [below,gray]{};
\draw[gray] (-1.89+0.25,-1.8+0.05) -- (-1.89+0.25,-1.8-0.05) node [below]{};
\draw[gray] (-1.89+0.6,-1.8+0.05) -- (-1.89+0.6,-1.8-0.05) node [below ]{};
\draw[gray] (-1.89+0.8,-1.8+0.05) -- (-1.89+0.8,-1.8-0.05) node [below ]{};
\draw[gray] (-0.89,-1.8+0.05) -- (-0.89,-1.8-0.05) node [below]{};
\useasboundingbox[black, font=\tiny] (-1.8,-1.8-0.3) -- (-0.6,-1.8-0.3) node[pos=0.14,rotate=-30]{$R_C$} node[pos=0.45,rotate=-30]{$t_{AlSb}$} node[pos=0.75,rotate=-30]{$t_{InAs}$}  node[pos=0.95,rotate=-30]{$t_{GaSb}$};
\end{tikzpicture}

\caption{(a) Band diagram for the InAs/GaSb/AlSb material system. Sketches showing (b) the quantum well structure and (c) the core-shell-shell nanowire we consider.}\label{fig_sketch}
\label{fig_lines}
\end{figure}

%bulk
The InAs/GaSb material system has attracted interest due to its broken band gap alignment, see Fig.~\ref{fig_sketch}(a), with large overlap of conduction bands (CBs) and valence bands (VBs), leading to hybridized electron-hole states in low-dimensional systems. %\cite{some papers} QWs 
The system has previously been studied in quantum wells (QWs),\cite{Liu2008,Knez2011,Nichele2016edge} often sandwiched between AlSb barriers, see Fig.~\ref{fig_sketch}(b). 
Compared to the broken band gap alignment in bulk, confinement in the QW moves the CBs up and the VBs down, which can restore a band gap.  However, if confinement is not large enough to give a conventional band gap, hybridization of the CBs and VBs can still cause an effective band gap to open up. We define such a hybridization gap as a band gap where the VB band edge lies above the CB band edge.
%In QWs an effective band gap opens up due to confinement and hybridization between CB- and VB-like states. When confinement is dominating a normal confinement gap is present, while large hybridization gives rise to an inverted, hybridization gap.
InAs/GaSb QWs are known for exhibiting the quantum spin Hall effect in this inverted regime,\cite{Liu2008,Knez2011} hence being topological insulators.\cite{Hasan2010}  
The topological insulators host edge states which are spin-momentum locked, carrying spin currents in two opposite directions. These states are robust to perturbations as long as time reversal symmetry is not broken. %The QSH effect was first predicted for graphene \cite{KaneMele2005a,KaneMele2005b}, and later for HgTe/CdTe quantum wells \cite{BHZ2006} where it was first discovered experimentally in 2007 \cite{König2007}. %Quantum wells made of InAs/GaSb are another two-dimensional (2D) material system predicted to be a topological insulator \cite{Liu2008} that has been verified experimentally\cite{Knez2011}. 

%NWs
In addition to the QWs, InAs/GaSb core-shell nanowires (NWs) have been investigated both experimentally and theoretically.\cite{Vinas2016,Kishore2012,Luo2016SR,Ek2011,Ganjipour2012,Ganjipour2015,Gluschke2015,Rieger2015,Namazi2015,Nilsson2016,Rocci2016}  Core-shell NWs with one shell are grown by several groups today,\cite{Lauhon2002,Ek2011,Ganjipour2012,Ganjipour2015,Gluschke2015,Rieger2015,Namazi2015,Nilsson2016,Rocci2016} and NWs with two shells can be grown,\cite{Lauhon2002} e.g., for the reason of a passivating outer layer on InAs.\cite{Nilsson2016}%\cite{moregrowth}

%Our work
In this work we study core-shell-shell NWs, where an insulator core is radially overgrown with one InAs and one GaSb layer, see Fig.~\ref{fig_sketch}(c). 
This system is in class AII which lacks a topological phase in one dimension (1D).\cite{Altland1997} However, a core-shell-shell NW, as depicted in Fig.~\ref{fig_sketch}(c), is not strictly 1D (and in the limit of an infinite radius it tends to a 2D QW system). %We consider such a core-shell-shell NW, made of AlSb, InAs and GaSb, and investigate its spectrum and wave functions. 
Using the $\mathbf{k \cdot p}$ Kane model we show that the hybridization gap seen in the QW persists in the NW we consider, for suitable InAs and GaSb shell thicknesses. Using a Bernevig-Hughes-Zhang (BHZ) model, \cite{BHZ2006} with parameters taken from fitting to the \kp band structures, we study a finite NW, and conclude that the core-shell-shell NWs can host end states. However, even though these end states originate from the QW edge states, they are different, since the edge states gap out when we ``roll up" the QWs. The NW end states are not robust to axial disorder, in contrast to the topologically protected edge modes in QWs. However, the end states are robust to (time-reversal preserving) disorder in the angular direction of the NW. %, but are only subject to an energy shift.

%Organzation
The paper is organized as follows: in section II we present the Kane model, the BHZ model and the tight-binding (TB) scheme we use to calculate the spectra and the wave functions of the NW and QW systems. In section III we present the results together with a discussion. Section IV contains a brief conclusion. %  We conclude our findings in section IV. 
  
%%%%%%%%%%%%%%%%%%%%%%%%%%%%%%%%%%%%%%%%%%%%%%
%%%%%%%%%%%%%%%%%%%%%%%%%%%%%%%%%%%%%%%%%%%%%%
%%%%%%%%%%%%%%%%%  Figure 1 %%%%%%%%%%%%%%%%%%%%%%%
%%%%%%%%%%%%%%%%%%%%%%%%%%%%%%%%%%%%%%%%%%%%%%
%%%%%%%%%%%%%%%%%%%%%%%%%%%%%%%%%%%%%%%%%%%%%%

%%%%%%%%%%%%%%%%%%%%%%%%%%%%%%%%%%%%%%%%%%%%%%
%%%%%%%%%%%%%%%%%%%%%%%%%%%%%%%%%%%%%%%%%%%%%%
%%%%%%%%%%%%%%%%%  Method %%%%%%%%%%%%%%%%%%%%%%%
%%%%%%%%%%%%%%%%%%%%%%%%%%%%%%%%%%%%%%%%%%%%%%
%%%%%%%%%%%%%%%%%%%%%%%%%%%%%%%%%%%%%%%%%%%%%%

\section{Method}

We use a Kane model \cite{Kane1957} to obtain the band structures of the QWs and the NWs. The Kane Hamiltonian is given by\cite{Kane1957,Foreman1997,Birner2011} 
\begin{equation}
H = 
\begin{pmatrix} 
H_{4 \times 4}  & 0 \\
0 & H_{4 \times 4}
\end{pmatrix} 
+ H_{SOC}, \label{eq_kane}
\end{equation}
with 
\begin{widetext}
\begin{equation}
\begin{split}
H_{4 \times 4} = &\eps_v(\mathbf{k}) \begin{pmatrix}
0 & 0 \\
0 & \mathds{1}_{3 \times 3}
\end{pmatrix} \\
&+\begin{pmatrix}
H_{CC} & iPk_x & iPk_y & iPk_z \\
\dagger &  k_xLk_x + k_yMk_y+ k_zMk_z & k_x\frac{N}{2}k_y + k_y\frac{N}{2}k_x &  k_x\frac{N}{2}k_z + k_z\frac{N}{2}k_x \\
\dagger & \dagger  &   k_xMk_x + k_yLk_y+ k_zMk_z & k_y\frac{N}{2}k_z + k_z\frac{N}{2}k_y \\
\dagger & \dagger & \dagger & k_xMk_x + k_yMk_y+ k_zLk_z
\end{pmatrix},
\end{split}
\end{equation}
\end{widetext}
with
\begin{equation}
H_{CC} = E_c + k_xAk_x + k_yAk_y + k_zAk_z,
\end{equation}
\begin{equation}
\eps_v(\mathbf{k})= E_v-\frac{\Delta_{SOC}}{3}+ \frac{\hbar^2}{2m_0}\mathbf{k}^2
\end{equation}
and
\begin{equation}
H_{SOC} = \frac{\Delta_{SOC}}{3} \begin{pmatrix}
0	&	0	&	0	&	0	&	0	&	0	&	0	&	0 \\
0	&	0	&	-i	&	0	&	0	&	0	&	0	&	1 \\
0	&	i	&	0	&	0	&	0	&	0	&	0	&	-i \\
0	&	0	&	0	&	0	&	0	&	-1	&	i	&	0 \\
0	&	0	&	0	&	0	&	0	&	0	&	0	&	0 \\
0	&	0	&	0	&	-1	&	0	&	0	&	i	&	0 \\
0	&	0	&	0	&	-i	&	0	&	-i	&	0	&	0 \\
0	&	1	&	i	&	0	&	0	&	0	&	0	&	0
\end{pmatrix}.
\end{equation}
Here the CB and VB band-edge energies are given by $E_c$ and $E_v$, respectively, so that the bulk band gap is $E_g = E_c - E_v$.
The Hamiltonian is written in the CB - VB and spin basis
\begin{equation}
\begin{split}
\{ &\ket{S\uparrow}, \ket{P_X\uparrow}, \ket{P_Y\uparrow}, \ket{P_Z\uparrow},  \\
&\ket{S\downarrow},  \ket{P_X\downarrow}, \ket{P_Y\downarrow}, \ket{P_Z\downarrow}\},
\end{split}
\end{equation}
where the CB states $\{ \ket{S} \}$ are given by $s$ orbitals and the VB states $\{ \ket{P} \}$ are given by $p$ orbitals.
The parameters used in the Kane Hamiltonian are given in terms of Luttinger parameters and the electron vacuum mass $m_0$ as 
\begin{equation}
\begin{aligned}
P & = \sqrt{\frac{\hbar^2}{2m_0}E_P } \\
L &= -\frac{\hbar^2}{2m_0}(\gamma_1+4\gamma_2) + \frac{P^2}{E_g} \\
M &= -\frac{\hbar^2}{2m_0}(\gamma_1 - 2\gamma_2)  \\
N &=  -6\frac{\hbar^2}{2m_0}\gamma_3 + \frac{P^2}{E_g} \\
A &= \frac{\hbar^2}{2m_0}.
\end{aligned}
\end{equation}
To be able to compare to works on the InAs/GaSb QWs, we use the parameters from Ref.~\onlinecite{Halvorsen2000}, in accordance with Refs.~\onlinecite{Zakharova2001,Nichele2016}. The parameter values are given in Table~\ref{tab_kp}.
\begin{table}
\caption{\label{tab_kp}The parameters used in the Kane model. The VB offsets between the materials are given by $E_{GaSb/InAs} = 0.56$~eV and $E_{AlSb/InAs} = 0.18$~eV.}
\begin{tabular}{llll}\toprule
  & InAs & GaSb & AlSb \\
  \colrule
  $E_g$ (eV) & 0.41&  0.8128& 2.32\\
  $E_P$ (eV) & 22.2 & 22.4& 18.7 \\
  $\gamma_1$  & 19.67& 11.80& 4.15\\
  $\gamma_2$  & 8.37& 4.03& 1.01\\
  $\gamma_3$  & 9.29& 5.26& 1.75\\
  %    $A$  & & & \\
  $\Delta_{SOC}$ (eV) & 0.38 & 0.752 & 0.75\\
  %  $m_c^*/m_0$  & & & \\
\botrule
\end{tabular}
\end{table}
We consider the QWs and the NWs to be grown in the $[111]$ direction. To obtain the Kane Hamiltonian in this crystallographic direction, we must impose a rotation of the coordinate system of the Hamiltonian\cite{Willatzen2009}. This process is discussed in Ref.~\onlinecite{Vinas2016} and follows Refs.~\onlinecite{Lassen2006,Willatzen2009,NWLuo}.
We solve the Schr\"odinger equation for the Kane Hamiltonian within the envelope function approximation, to include the effect of the different materials and geometries.  The envelope function approximation is employed by substituting  $k_n \rightarrow -i \partial_n $ in Eq.~(\ref{eq_kane}) for the directions $n$ where translational symmetry is broken. We then use a basis function expansion of the envelope functions $\psi$. In the calculations for the QW, a plane wave basis is used in the growth direction $z$.
%\begin{equation}
%\psi(x,y,z) = \sum_{n=1}^{N\rightarrow\infty} \eta(n) e^{in\pi %z/L_z}e^{ik_x x}e^{ik_y y} %\frac{1}{\sqrt{L_z}}
%\end{equation} 
%where $z$ is considered the growth direction of the QWs and $\eta$ is a normalization constant. 
In the calculations for the NWs, assumed to be cylindrical, we assume plane waves in the growth direction $z$ and expand $\psi$ in a basis consisting of approximations to the Bessel functions far from the origin in the radial direction\cite{Abramowitz1974} $r$
\begin{equation}
f_{m,n}(r, \theta) = \eta(m,n) \frac{1}{\sqrt{r}}\sin \left(\frac{R-r}{R_c-R} m\pi \right)e^{in\theta},% e^{ik_z z}.  %\frac{2}{\sqrt{\pi}}
\end{equation}
where $\eta(m,n)$ is a normalization factor.
%For this case $z$ is considered the growth direction of the NW.
In the calculations the basis expansions are truncated after convergence is reached.

 %(should cite additional works here?)
%We choose this configuration because this is what makes sense growth-wise (both in 1D, where it's probably hard to have a hollow core, and in 2D where they, as far as I know, always have the InAs/GaSb sandwiched between AlSb on both sides). then we can also comment on the spurious solutions

In the NW system we use one inner AlSb barrier, as in Fig.~\ref{fig_sketch}(c), while for the QW calculations we use AlSb barriers on both sides of the structure (see Fig.~\ref{fig_sketch}b). We choose these configurations, because this is how the structures would most likely be grown. In addition, these configurations avoid problems with spurious solutions. The spurious solutions are unphysical solutions to the Schr\"odinger equation for the Kane Hamiltonian that can arise when employing the envelope function approximation.\cite{Foreman1997,Winkler2003,Birner2011,Willatzen2009} For computational reasons, we consider the core-shell-shell NW to be hollow, in the sense that the inner core consists of vacuum, as in Fig.~\ref{fig_sketch}(c). 
However, in a real experimental structure the full core could be filled with AlSb, without affecting the energy dispersion for the states around the gap. One can also imagine another insulator core (or possibly vacuum) instead of this filled AlSb core. In this case, we expect that the results will not change qualitatively, because we see that the wave functions around the gap only penetrate very little into the AlSb layer.
%(or another insulator, such as vacuum), likely without any considerable changes of the results.
%and we wish to avoid expensive computational costs. 

We use a BHZ and a TB model together with the \kp calculations to study end states of a finite NW and to add disorder to the system. The reason that we use this model to study a finite system is that it becomes too numerically expensive to solve using our \kp model.
The BHZ model  \cite{BHZ2006,PhysRevLett.100.236601,Rothe2010} is given by
\begin{equation}
\label{TBhamtotal}
H = H_{BHZ} + H_{SIA},
\end{equation}
where
\begin{equation}
H_{BHZ} =   \begin{pmatrix}
	h(k)	&	0	 \\
	0	&	h^*(-k)	
\end{pmatrix} , 
\end{equation}
with
\begin{equation}
	h(k) =   \begin{pmatrix}
		\epsilon(k) + M(k)	&	Ak_+	 \\
		Ak_-	&	\epsilon(k) - M(k)	
	\end{pmatrix} , 
	\end{equation}
and $\epsilon(k) =  C - D (k_x^2 + k_y^2)$, $ M(k) = M-B(k_x^2 + k_y^2) $, $k_\pm = k_x \pm i k_y$. $H_{BHZ}$ is written in the basis $(\ket{CB+}, \ket{VB+},\ket{CB-},\ket{VB-})^T$. $\ket{CB\pm}$ ($\ket{VB\pm}$) corresponds to the lowest (highest) energy CB (VB) in InAs (GaSb) and $\pm$ denotes Kramers partners. Structural inversion asymmetry (SIA) spin splitting, which is intrinsic in the \kp model, is explicitly included here \cite{PhysRevLett.100.236601}:
\begin{equation}
\label{TBhamsia}
H_{SIA} =   \begin{pmatrix}
 0	         &	 0             &  -i R_0 k_-   & 0	 \\
 0           &	 0	           &   0           &  i T_0 k_-^3  \\
 i R_0 k_+   &   0             &  0            &0   \\
 0           &  - i T_0 k_+^3   & 0            &0
\end{pmatrix}.
\end{equation}
Along with the linear-$k$ ($R_0$) term coupling CB-like states, a cubic-$k$ ($T_0$) term coupling VB-like states has also been included, following Ref. \onlinecite{Rothe2010}. A second order SIA term coupling CB-like and VB-like states is also generally present, but it has very little effect on the band structures we wish to reproduce and is therefore omitted. We do not consider bulk inversion asymmetry terms, as they are negligible for InAs/GaSb QWs \cite{Molenkamp2013}.
%It would be nice to have a reason for why this is the case, so that we can set them to zero from the beginning.

The Hamiltonian in Eq.~(\ref{TBhamtotal}) can  be readily used to reproduce the \kp band structure in the infinite QW system. 
%by going to position space ($k_x \rightarrow -i\partial_x$, $k_y \rightarrow -i\partial_y$) and calculating its action on the discretized  positional states $\ket{x,y} = \ket{an_x,an_y} \equiv \ket{n_x,n_y}$, where $a$ is the lattice constant and $n_x$ $(n_y)$ is the site number in the $x$ $(y)$ direction. 
The finite QW can be studied by discretizing Eq.~(\ref{TBhamtotal}) on a square lattice of finite dimensions, and is found to host mid-gap Kramers-degenerate topological edge states for a wide parameter range.\cite{PhysRevLett.100.236601} Considering the 2D QW in the $y-z$ plane instead, and with periodic boundary conditions along the $y$ direction, the system is in a ``rolled up" geometry equivalent to a cylindrical NW with the growth axis along $z$. % For a NW infinite in $z$, the quasi-1D band structure is extracted for comparison with the \kp results. %For a finite NW, we find that the system hosts localized end states at both NW's ends. We study the effect of different types disorder on the end states in order to check their robustness against TR preserving perturbations.

%For a NW finite in $y$ we investigate the effect of imposing the $x$-periodicity on the topological edge states of the finite QW system. The mid-gap states are now localized at the NW's ends (end states) and we study their robustness against angular (along $x$) and axial (along $y$) time reversal (TR) symmetry preserving disorder.

%A tight-binding model is then used to reproduce the band structures, and to calculate the wave functions in a finite-length NW. %Thanos will here write a model description of the with the BHZ Hamiltonian.

%%%%%%%%%%%%%%%%%%%%%%%%%%%%%%%%%%%%%%%%%%%%%%
%%%%%%%%%%%%%%%%%%%%%%%%%%%%%%%%%%%%%%%%%%%%%%
%%%%%%%%%%%%%%%%%  Results %%%%%%%%%%%%%%%%%%%%%%%
%%%%%%%%%%%%%%%%%%%%%%%%%%%%%%%%%%%%%%%%%%%%%%
%%%%%%%%%%%%%%%%%%%%%%%%%%%%%%%%%%%%%%%%%%%%%%

\section{Results} 

%%%%%%%%%%%%%%%%%%%%%%%%%%%%%%%%%%%%%%%%%%%%%%
%%%%%%%%%%%%%%%%%%%%%%%%%%%%%%%%%%%%%%%%%%%%%%
%%%%%%%%%%%%%%%%%  Figure2 %%%%%%%%%%%%%%%%%%%%%%%
%%%%%%%%%%%%%%%%%%%%%%%%%%%%%%%%%%%%%%%%%%%%%%
%%%%%%%%%%%%%%%%%%%%%%%%%%%%%%%%%%%%%%%%%%%%%%

\begin{figure}
	\includegraphics[width=0.23\textwidth,trim={0cm 0cm 0cm 0cm},clip]{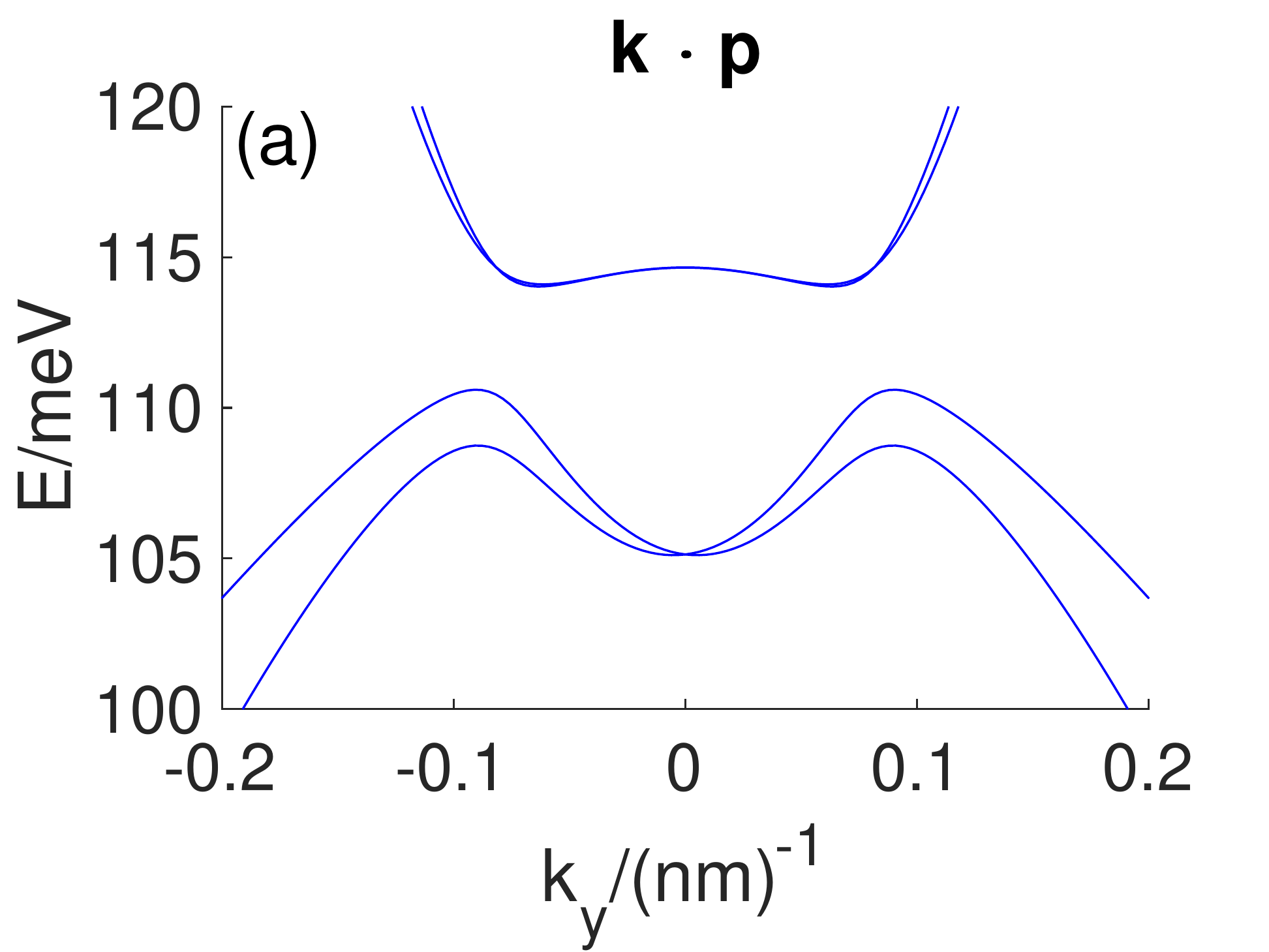}
	\includegraphics[width=0.23\textwidth,trim={0cm 0cm 0cm 0cm},clip]{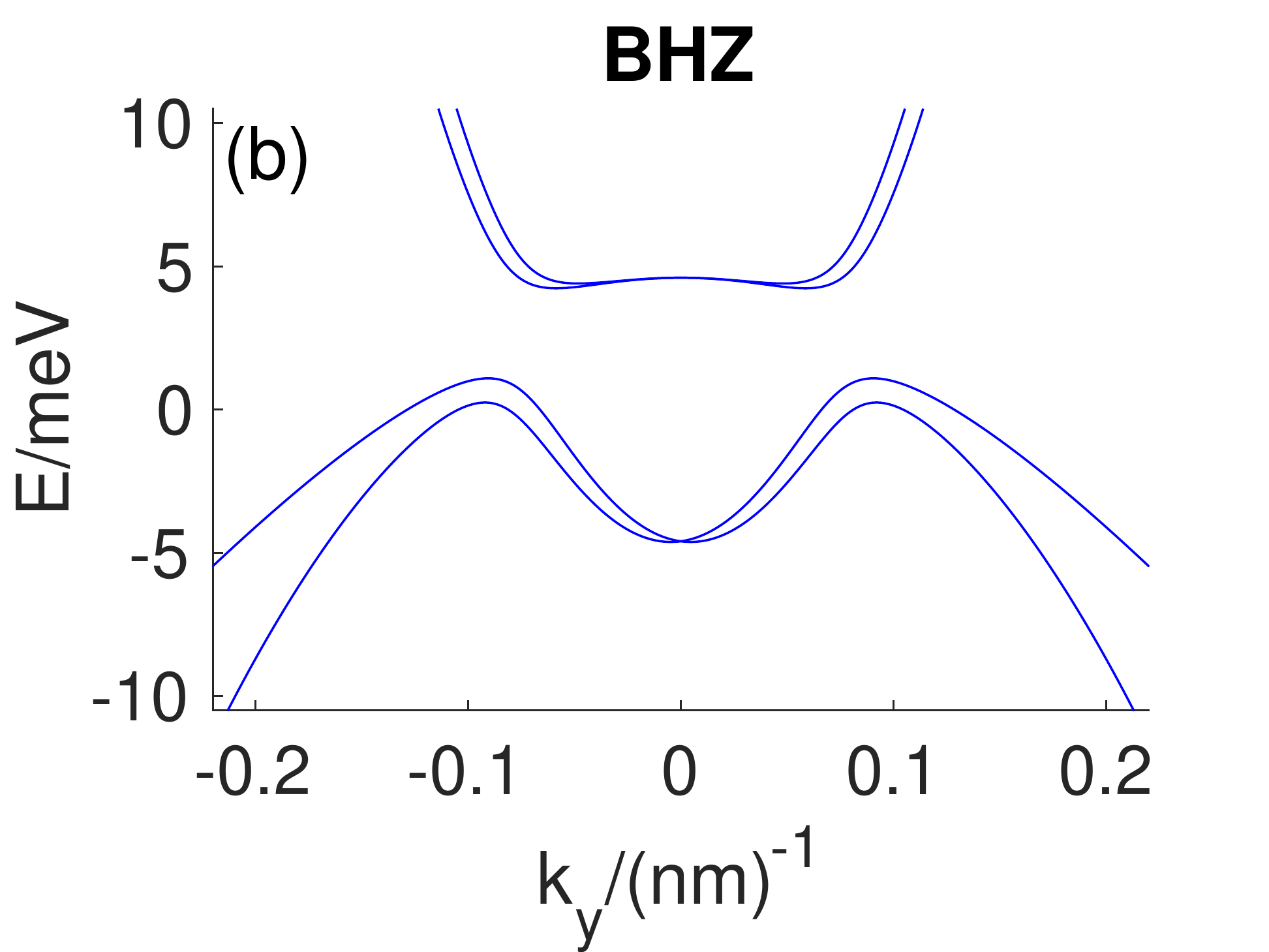}
	\caption{ (a)~Band structure for a QW calculated using \kp theory with $t_{AlSb} = 10$~nm, $t_{InAs} = 11$~nm and $t_{GaSb} = 5$~nm, for $k_x = 0$. (b)~Band structure reproduced with the BHZ Hamiltonian for an infinite QW system for $k_x = 0$. The used parameters are given in Table~\ref{BHZparam}.}
	\label{fig_2Dbands}
\end{figure}
%%%%%%%%%%%%%%%%%%%%%%%%%%%%%%%%%%%%%%%%%%%%%%
%%%%%%%%%%%%%%%%%%%%%%%%%%%%%%%%%%%%%%%%%%%%%%
%%%%%%%%%%%%%% BHZ parameters table %%%%%%%%%%%%%%%%%%%
%%%%%%%%%%%%%%%%%%%%%%%%%%%%%%%%%%%%%%%%%%%%%%
%%%%%%%%%%%%%%%%%%%%%%%%%%%%%%%%%%%%%%%%%%%%%%
%%%%%%%%%%% Attempt for full page table, doesn't look good
%\begin{table*}  % will be placed at top of a page
%   \begin{tabularx}{0.8\textwidth} { 
%		| >{\centering\arraybackslash}X 
%		| >{\centering\arraybackslash}X 
%		| >{\centering\arraybackslash}X
%		| >{\centering\arraybackslash}X
%		| >{\centering\arraybackslash}X 
%		| >{\centering\arraybackslash}X| }
%		\hline
%	$A(meV\cdot nm)$ & $B(meV\cdot nm^2)$  & $D(meV\cdot nm^2)$ & 		$M(meV)$ & $R_0(meV\cdot nm)$ & $T_0(meV\cdot nm^3)$  \\
%		\hline
%    	30.5  & -710  & -450 & -4.6 & 10  & 300  \\
%    	\hline
%    \end{tabularx}
%\caption{GC parameters on}
%\label{BHZparam}
%\end{table*}
\begin{table}  
	\begin{tabularx}{0.30\textwidth} { 
			| >{\centering\arraybackslash}X 
			| >{\centering\arraybackslash}X| }
		\hline
		$A$~(meV$\cdot$ nm) & 30.5    \\
		\hline
		$B$~(meV$\cdot$ nm$^2$)  & -710    \\
		\hline
		$D$~(meV$\cdot$ nm$^2$)  & -450     \\
		\hline
		$M$~(meV)  & -4.6     \\
		\hline
		$R_0$~(meV$\cdot$ nm) & 10    \\
		\hline
		$T_0$~(meV$\cdot$ nm$^3$)  & 300    \\
		\hline
	\end{tabularx}
	\caption{BHZ parameters used in Fig.~\ref{fig_2Dbands}(b). The rest of the parameters appearing in Eqs.~(\ref{TBhamtotal}-\ref{TBhamsia}) are set to zero. }
	\label{BHZparam}
\end{table}
%%%%%%%%%%%%%%%%%%%%%%%%%%%%%%%%%%%%%%%%%%%%%%
%%%%%%%%%%%%%%%%%%%%%%%%%%%%%%%%%%%%%%%%%%%%%%
%%%%%%%%%%%%%%%%%  Figure3  %%%%%%%%%%%%%%%%%%%%%%%
%%%%%%%%%%%%%%%%%%%%%%%%%%%%%%%%%%%%%%%%%%%%%%
%%%%%%%%%%%%%%%%%%%%%%%%%%%%%%%%%%%%%%%%%%%%%%
\begin{figure*}
\includegraphics[width=0.3\textwidth,trim={0cm 0cm 1cm 0cm},clip]{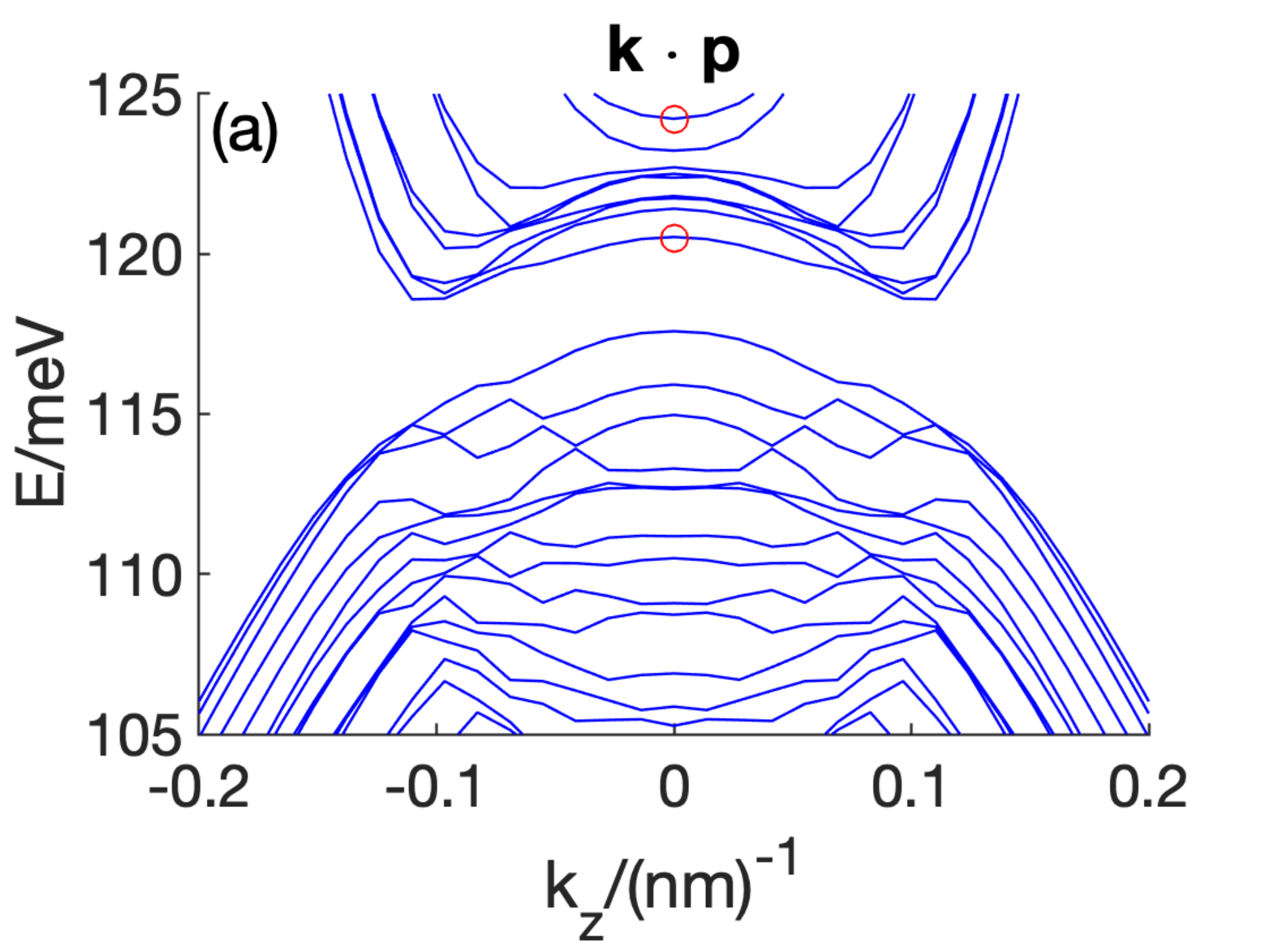}
\includegraphics[width=0.3\textwidth,trim={0cm 0cm 1cm 0cm},clip]{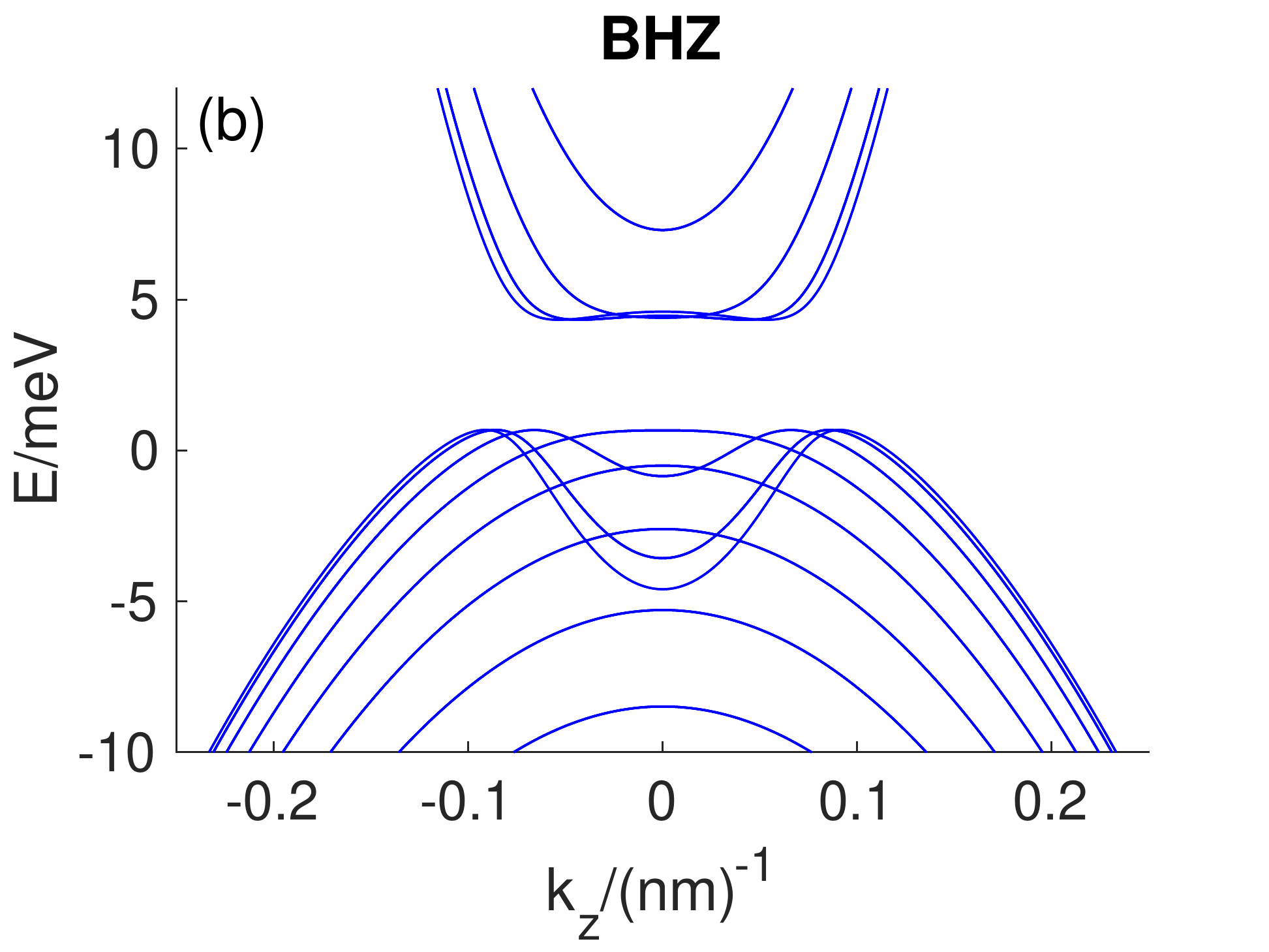}
\includegraphics[width=0.3\textwidth,trim={0cm 0cm 1cm 0cm},clip]{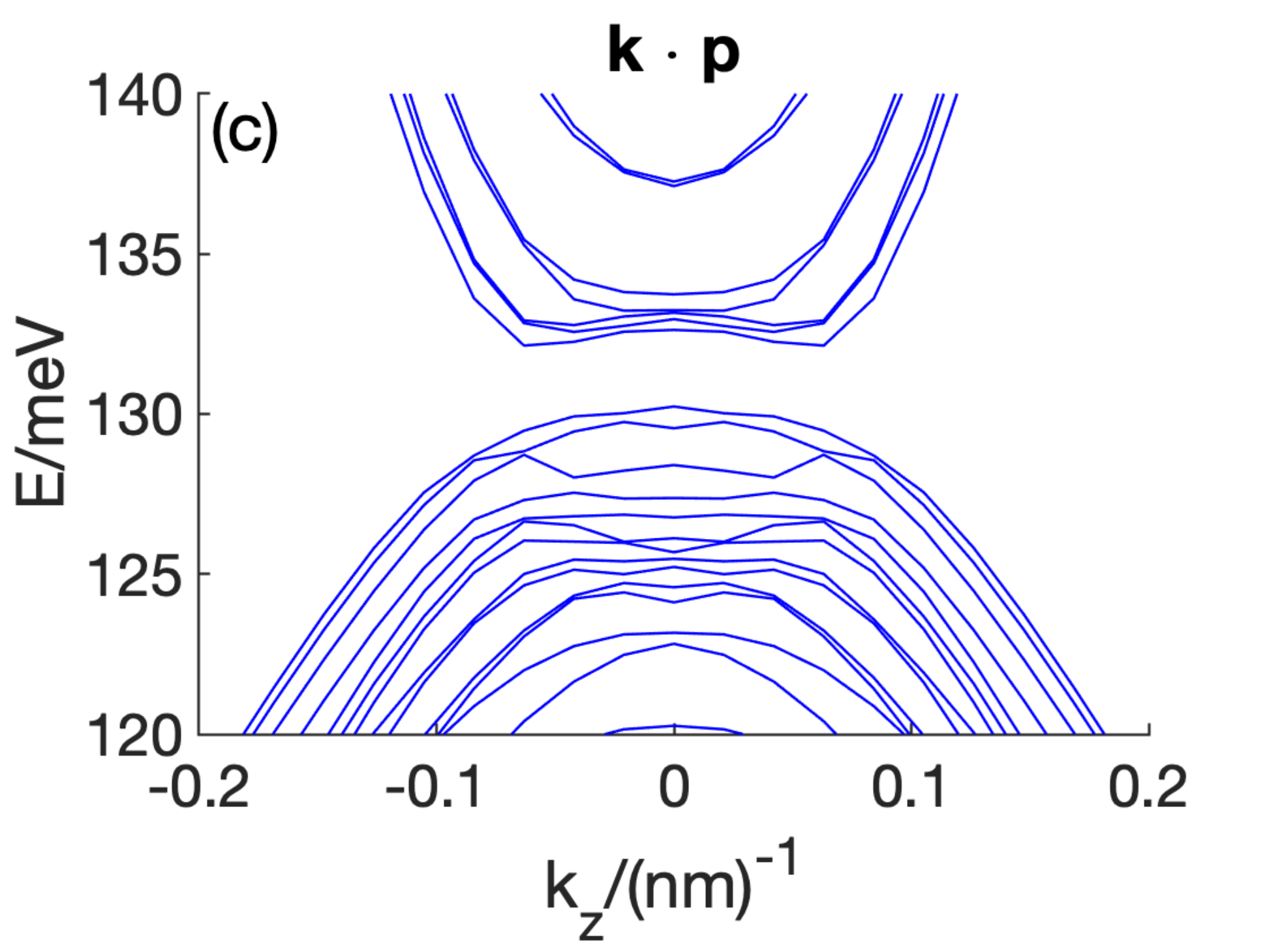}
\caption{ (a) Band structures calculated using the \kp model with $R_c = 10$~nm, $t_{AlSb} = 14$~nm, $t_{InAs} = 11$~nm and $t_{GaSb} = 5$~nm. (b) Band structure calculated using the BHZ model with $R_{BHZ} = 32$~nm and no SIA terms included ($R_0=T_0=0$). The rest of the parameters are the same as in Fig.~\ref{fig_2Dbands}(b). (c) Same as (a) but with $t_{InAs} = 9.5$~nm and $t_{GaSb} = 6.5$~nm .}
\label{fig_1Dbands}
\end{figure*}
Figure~\ref{fig_2Dbands}(a) shows the band structure for a QW with $t_{InAs} = 11$~nm and $t_{GaSb} = 5$~nm, calculated using \kp theory. The corresponding band structure from BHZ calculations are shown in Fig.~\ref{fig_2Dbands}(b). The parameters (see Table~\ref{BHZparam}) for the BHZ model are chosen to give a good match with the \kp band structure. The band structures are in good agreement with previous works\cite{Zakharova2001} and show a hybridization gap. For an ordinary confinement gap to open up, the gap needs to close and reopen. The QW is known to host topologically protected edge states in this inverted regime. %Adding disorder to the BHZ model, these states do not gap out. In Figure~\ref{fig_2Dbands}c) we present a map of the band gap size as a function of the InAs and GaSb thicknesses. Note that gap sizes bigger than the maximum hybridization gap is shifted down in energy, so that the gap closing between the topological and the trivial regime is clearly visible (the confinement gap can be tuned to arbitrarily large size if the materials become thin enough).

We now shift our attention to the main subject of our study, namely InAs/GaSb core-shell-shell NWs.
Figure~\ref{fig_1Dbands}(a) shows the band structure for such a core-shell-shell NW, calculated using the Kane model, with $t_{InAs} = 11$~nm, $t_{GaSb} = 5$~nm, $R_C = 10$~nm and $t_{AlSb} = 14$~nm, so the thicknesses of InAs and GaSb are the same as in Fig.~\ref{fig_2Dbands}(a). The band gap is only $E_g = 1$~meV, compared to the larger value of $E_g = 3.4$~meV for the QW (for the same thicknesses of InAs and GaSb). The main reason for this smaller gap is that the confinement effects are different in the NW system, and possibly that in the core-shell-shell NW the curvature effects also become important. 
The angular subbands are much closer in energy than the radial ones, which makes sense if we compare the length scales: the radial confinement is roughly $t_{InAs}+t_{GaSb} = 16$~nm, while confinement in the angular direction is of the magnitude $2\pi R \approx 250$~nm. All subbands are two-fold degenerate, because both time reversal and structural inversion symmetries are present.%There are three main effect that determines the hybridization gap size; the confinement in InAs, the confinement in GaSb, and the overlap of the materials. While in 2D, the bands and the band gap size is determined by $t_{InAs}$, $t_{GaSb}$ and $\frac{t_{InAs}}{t_{GaSb}}$, in 1D the curvature (or angular confinement) effects also become important. In a given geometry (such as the core-shell-shell NWs we consider), the three parameters cannot be tuned separately from each other ($t_{InAs}$ and $t_{GaSb}$ will affect all three properties). One approach to think about this is that the material distribution (i.e., the volume quota) between the two materials should be constant to give a similar gap size. However, this condition must be relaxed if we want to keep the confinement effects roughly  the same in each material respectively. Nevertheless, we do not exclude the possibility to find a larger hybridization gap in the core-shell-shell system by tuning the material thicknesses.

In Fig.~\ref{fig_1Dbands}(b) we show a band structure calculated with the BHZ model, using the same parameters as in Fig.~\ref{fig_2Dbands}(b), but with periodic boundary conditions along one direction. To reflect the symmetries of the cylindrical NW geometry, we set the SIA terms to zero ($R_0=0$, $T_0=0$). % Here, we consider the system to lie on the $y-z$ plane and take periodic boundary conditions along $y$, in order to have a NW with the same growth direction as in the \kp calculation (along $z$).
%The system is rotated so that the growth axis is now aligned with the $z$ axis in order to be consistent with the \kp coordinate system, a practice we adopt for the rest of the paper.
We choose the value of the radius ($R_{BHZ}=32$~nm) to be in between the inner and outer radii in the \kp calculations. We believe that the main reason that the band structures in  Fig.~\ref{fig_1Dbands}(a) and (b) look so different is that the confinement effects in the \kp calculations become very different in a cylindrical geometry. 
%However, by a reasonable adjustment of the shell thicknesses in the \kp calculations we can obtain a similar band structure to the one in Fig.~\ref{fig_1Dbands}(b).

Figure~\ref{fig_1Dbands}(c) shows that we can obtain a band structure from \kp calculations that is similar to the one in Fig.~\ref{fig_1Dbands}(b) by changing the shell thicknesses of the InAs and the GaSb shells to $t_{InAs} = 9.5$~nm and $t_{GaSb} = 6.5$~nm. The core radius and the AlSb shell thickness are $R_C = 10$~nm and $t_{AlSb} = 14$~nm, the same as in Fig.~\ref{fig_1Dbands}(a). We note that the hybridization gap is $E_g^{k \cdot p}=1.9$~meV, smaller than the hybridization gap $E_g^{BHZ}=3.7$~meV from the BHZ results. %The parameters in Fig.~\ref{fig_1Dbands}(c) are chosen to reflect the ``rolled up" QW in Fig.~\ref{fig_1Dbands}(b). 
\begin{figure}
    \centering
    \includegraphics[width=0.34\textwidth,trim={0cm 0cm 0cm 0cm},clip]{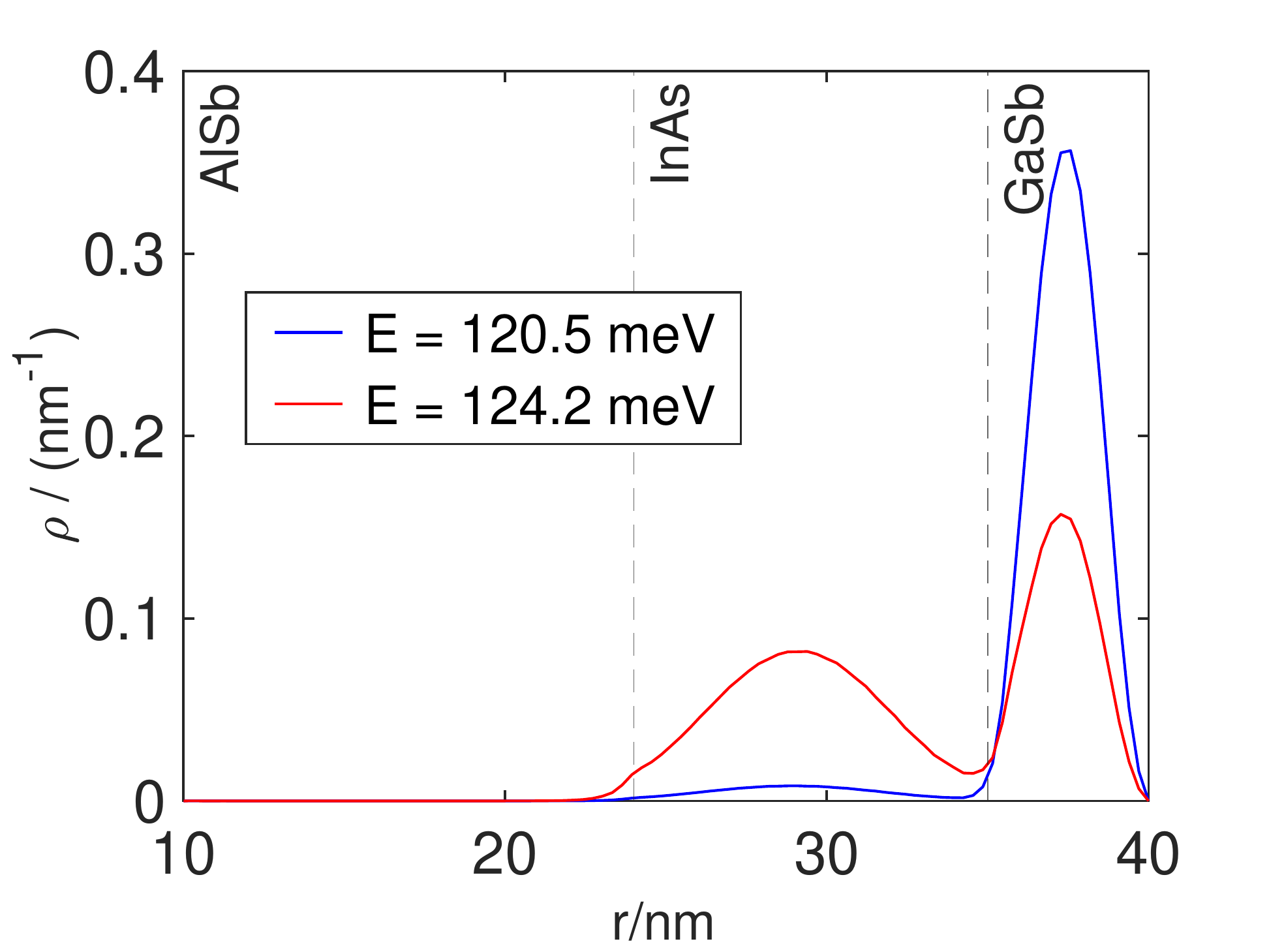}
    \caption{Probability density $\rho(r)$ for the two subbands marked with red in Fig.~\ref{fig_1Dbands}(a), for $k_z = 0$: the first subband above the gap (blue line) and the topmost subband seen in Fig.~\ref{fig_1Dbands}(a) (red line). }
    \label{fig_prob_dens}
\end{figure}
%Figure 4 shows the probability density in the radial direction [ekv.] for two different subbands $j$ in Fig. 3(a).
Figure~\ref{fig_prob_dens} shows the probability density in the radial direction $\rho_j(r) = r |\psi_{j,k_z=0}(r)|^2$ for two different subbands $j$ in Fig.~\ref{fig_1Dbands}(a). The probability density for the lowest-lying subband above the gap is plotted in blue and shows a state mostly confined in the outer shell. The red line shows the probability density for the topmost subband seen in Fig.~\ref{fig_1Dbands}(a). Even though this subband looks like a pure CB state for small $k_z$, we see that the state has large weight in both the InAs and the GaSb shell. In general, most of the subbands around the band gap have weight in both these outer shells, or predominantly in the GaSb shell. To find states confined in the InAs shell, one has to study subbands much higher ($\sim 50
$~meV) above the gap.% It is interesting that states close to the gap, that has a CB-like appearance in the band structure, have large weights in the GaSb shell. 
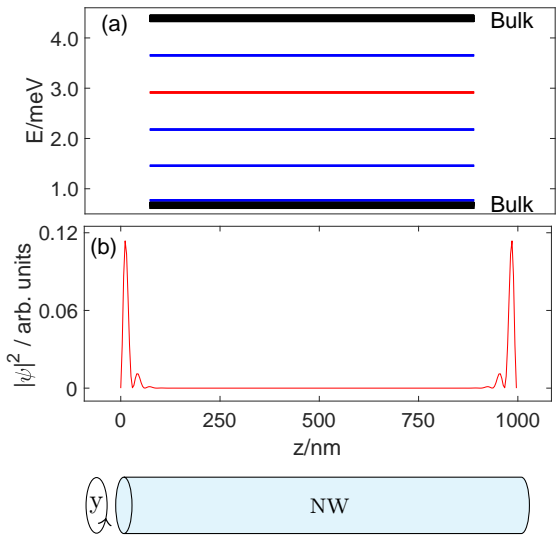
\begin{figure}
	\includegraphics[width=0.45\textwidth,trim={0cm 0.8cm 0cm 0cm},clip]{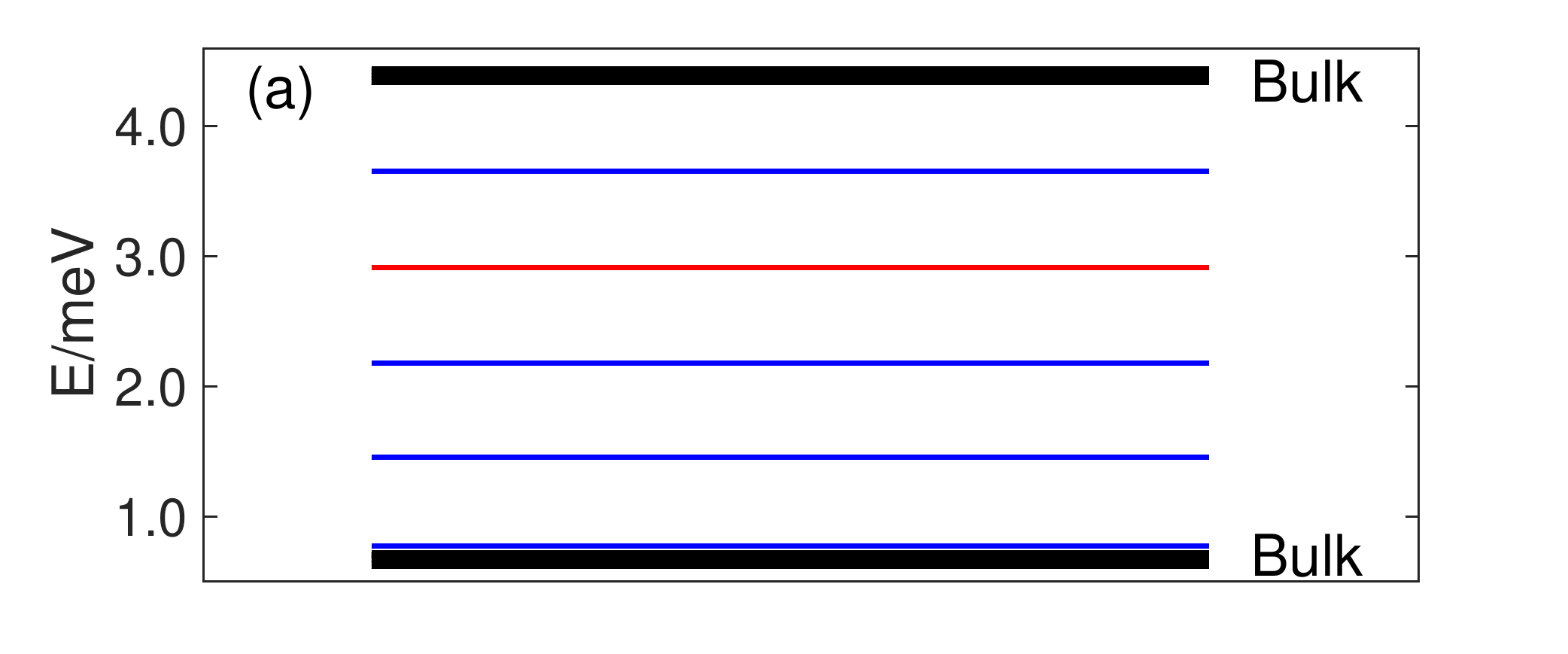}
	\hspace*{-0.1cm}\includegraphics[width=0.45\textwidth,trim={0cm -0.5cm 0cm 0.5cm},clip]{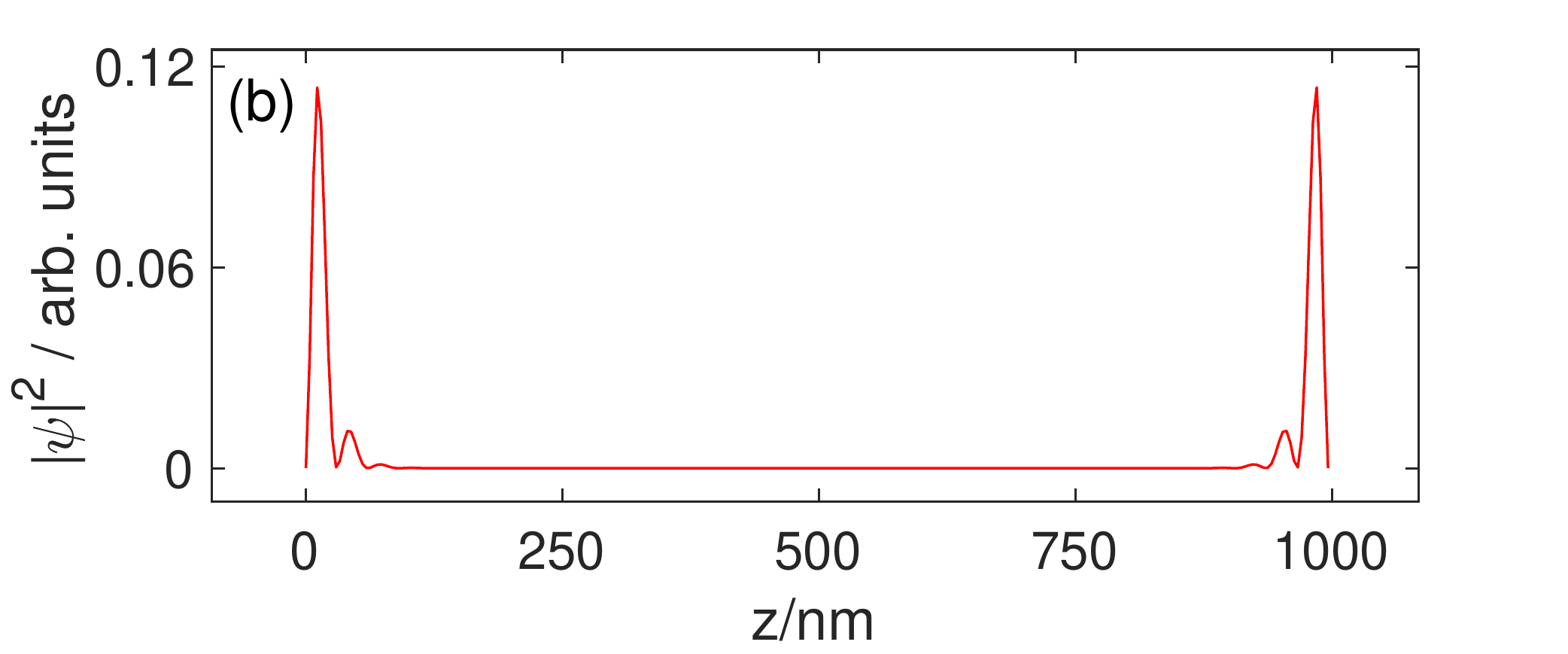}
	\hspace*{1.5cm}\begin{tikzpicture}[scale=1]
	\hspace*{-0.8cm}\node [cylinder, black, fill=cyan!10, rotate=180, draw, aspect=1.0,minimum height=55mm, minimum width=7.5mm] at (0,0) (c)  {};
	%\draw[black,->] (-1.5,-0.69) -- (2*1.89/3,-0.69) node [at end, right, black] {z};
	\node at (0,0) {\footnotesize NW}; 
	\draw[
	decoration={markings, mark=at position -0.1 with {\arrow[scale=1.5,black]{>}}    },
	postaction={decorate}]
	(-3.1,0) ellipse (0.14cm and 0.37cm);
	\node at (-3.1,0) {\normalsize y}; 
	%\node [cylinder, black, fill=cyan!10, rotate=180, draw, aspect=1.0,minimum height=55mm, minimum width=7.5mm] at (0,0) (c)  {};
	%\node at (0,0) {\footnotesize NW};   
	%\includegraphics[width=0.50\textwidth,trim={0cm -0.5cm 0cm 0cm},clip]{figs/5_disaxplot.pdf}
	%\hspace*{-0.1cm}\includegraphics[width=0.5\textwidth,trim={0cm -0.5cm 0cm 0.5cm},clip]{figs/5_gsmagndisplot.pdf}
	%\hspace*{0.5cm} \begin{tikzpicture}[scale=1]
	%\node [cylinder, black, fill=cyan!10, rotate=180, draw, aspect=1.0,minimum height=60mm, minimum width=7.5mm] at (0,0) (c)  {};
	%\draw[black,->] (-1.5,-0.69) -- (2*1.89/3,-0.69) node [at end, right, black] {z};
	%\node at (0,0) {\footnotesize NW}; 
	%\draw[,black,->] (-2.8,-0.69) -- (3.4,-0.69) node [at end, above, black] {y/nm};
	%\draw[black ] (-2.8,-0.69) -- (-2.8,-0.69 - 0.15) node [below, black] {0};
	%\draw[black ] (-2.8+1.35,-0.69) -- (-2.8+1.35,-0.69 - 0.15) node [below, black] {250};
	%\draw[black ] (-2.8+2.7,-0.69) -- (-2.8+2.7,-0.69 - 0.15) node [below, black] {500};
	%\draw[black ] (-2.8+4.05,-0.69) -- (-2.8+4.05,-0.69 - 0.15) node [below, black] {750};
	%\draw[black] (-2.8+5.4,-0.69) -- (-2.8+5.4,-0.69 - 0.15) node [below, black] {1000};
	%\draw[gray,<->] (-1.89,-1.8) -- (-1.89+0.4,-1.8) node [below, gray] {$R_C$};
	\end{tikzpicture}
	%\begin{tikzpicture}
	%\draw[black,thick]  -- (0,) node[draw]  {Placeholder};
	%\end{tikzpicture}
	\caption{(a) End and bulk state energies for a NW of length $L_{NW} = 1000$~nm obtained from BHZ/TB calculations. The zero angular momentum end state is colored red and the higher angular momentum end states blue, while the bulk states are black. The rest of the parameters are the same as in Fig.~\ref{fig_1Dbands}(b). (b) Probability density of the zero angular momentum wave function along the NW growth axis.}
	\label{fig_1Dbands_fin}
\end{figure}

Next, we use the BHZ model with the fitted parameters to study the fate of the QW's edge states in the NW geometry. This is meant in the sense that one can imagine arriving at a finite NW geometry by ``rolling up'' a finite 2D QW system. Coupling two of the edges causes the edge states to gap out. However, we find that this leaves localized end states at both NW ends. Figure~\ref{fig_1Dbands_fin}(a) shows the energies of the end states together with the more closely spaced bulk states using the same parameters as in Fig.~\ref{fig_1Dbands}(b), but the system is now taken to be finite in $z$ ($L_{NW} = 1000$~nm). One major difference between the end states seen for the NW system compared to the edge states in the QW, is that the NW end states are doubly degenerate \emph{in each Kramers sector} (4-fold degenerate in total), while for the QW the edge states are only Kramers degenerate. In Fig.~\ref{fig_1Dbands_fin}(b) the probability density along the NW for the zero angular momentum end state is plotted. For the chosen NW length the weights of the wave functions in this 4-fold degenerate subspace are highly localized at the NW ends.

\begin{figure}
	\includegraphics[width=0.23\textwidth,trim={0cm -0.50cm 0cm 0cm},clip]{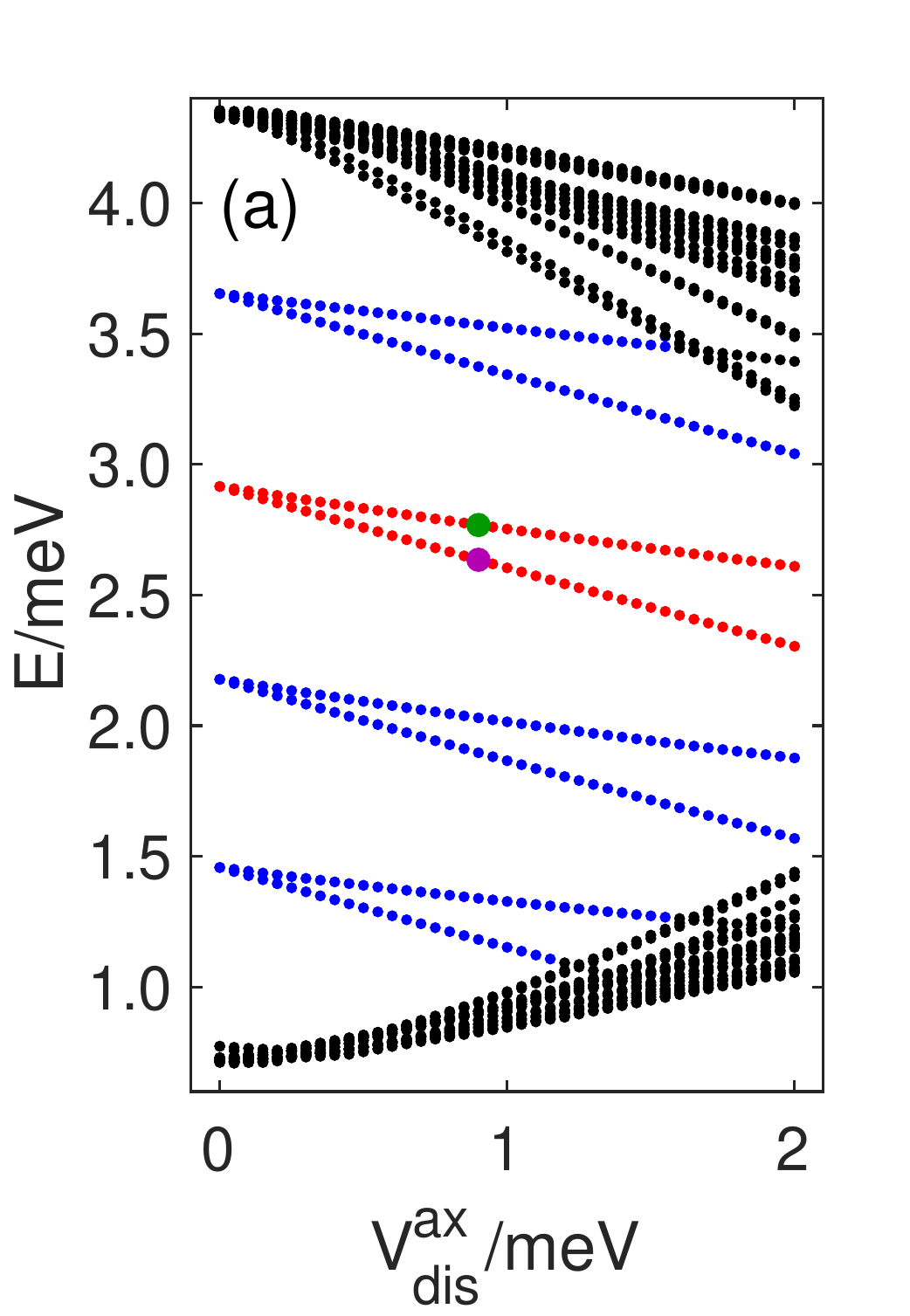}
	\includegraphics[width=0.23\textwidth,trim={0cm -0.50cm 0cm 0cm},clip]{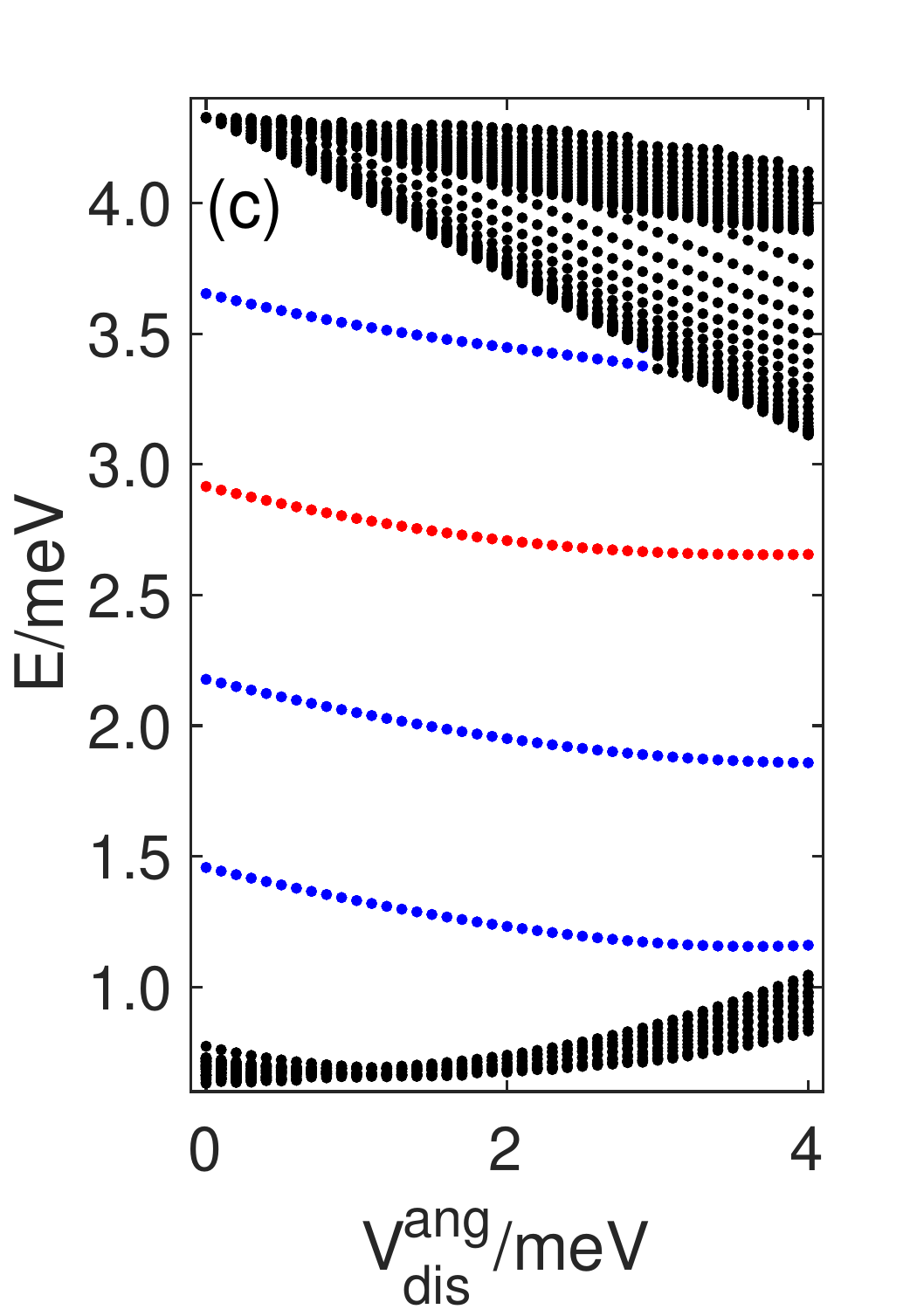}
	\hspace*{0.0cm}\includegraphics[width=0.5\textwidth,trim={0cm -0.5cm 0cm 0.5cm},clip]{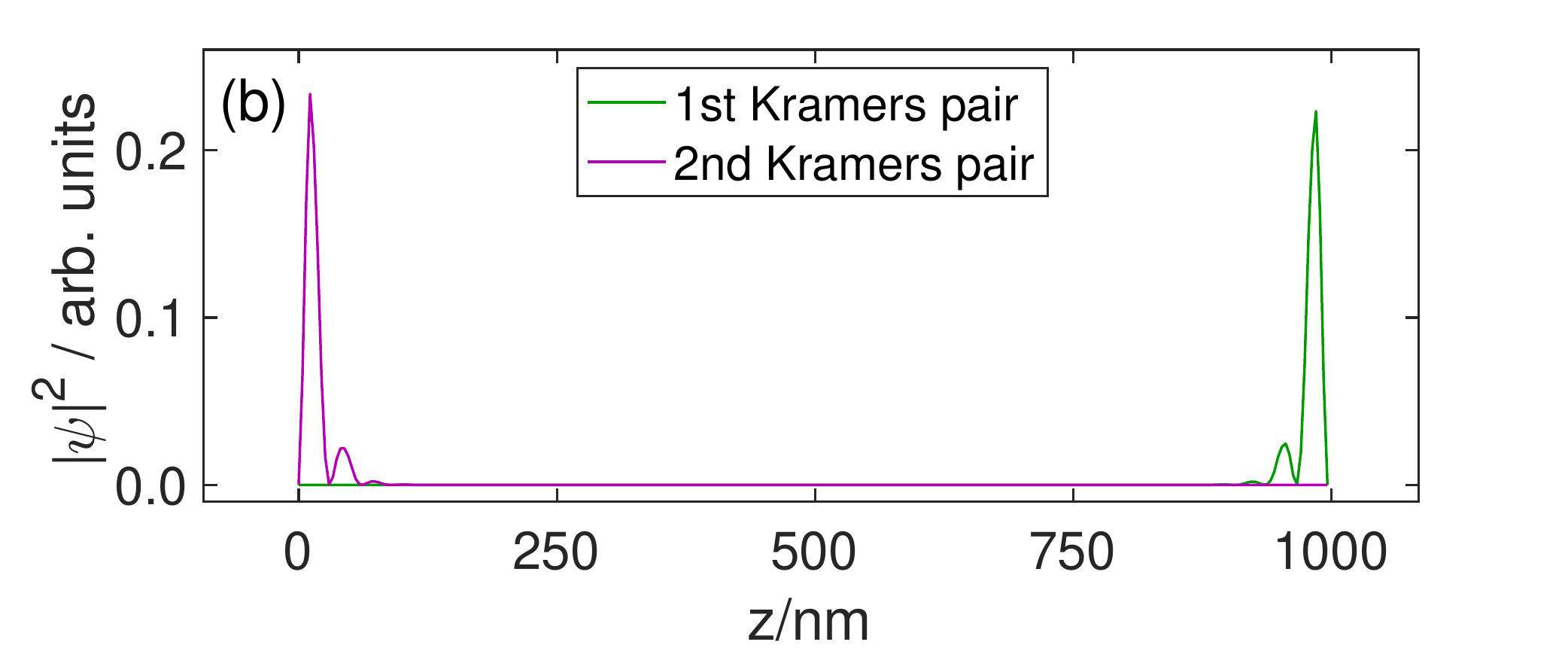}
	\hspace*{0.2cm} \begin{tikzpicture}[scale=1]
	\node [cylinder, black, fill=cyan!10, rotate=180, draw, aspect=1.0,minimum height=60mm, minimum width=7.5mm] at (0,0) (c)  {};
	%\draw[black,->] (-1.5,-0.69) -- (2*1.89/3,-0.69) node [at end, right, black] {z};
	\node at (0,0) {\footnotesize NW};
	\draw[
	decoration={markings, mark=at position -0.1 with {\arrow[scale=1.5,black]{>}}    },
	postaction={decorate}]
	(-3.4,0) ellipse (0.14cm and 0.37cm);
	\node at (-3.4,0) {\normalsize y};
	%\draw[fill=pink!40, font=\small] (-1.89,-0.69) circle (0.8)  node[anchor=south west] {};
	%\draw[fill=yellow!13, font=\small] (-1.89,-0.69) circle (0.6)  node[anchor=south west] {};
	%\draw[fill=white, font=\small] (-1.89,-0.69) circle (0.25)  node[anchor=south west] {};
	%\draw[gray,->] (-1.89,-0.69) -- (2*1.89/3,2*0.69/3) node [at end, right, gray] {z};
	%\draw[gray,->] (-1.89,-0.69) -- (-0.89,-0.69) node [at end, right, gray] {r};
	%\draw[gray,<->] (-1.89,-1.8) -- (-1.89+0.4,-1.8) node [below, gray] {$R_C$};
	%\draw[gray] (-1.89,-1.8) -- (-0.89,-1.8) node [below, gray] {};
	%\draw[gray] (-1.89,-1.8+0.05) -- (-1.89,-1.8-0.05) node [below,gray]{};
	%\draw[gray] (-1.89+0.25,-1.8+0.05) -- (-1.89+0.25,-1.8-0.05) node [below]{};
	%\draw[gray] (-1.89+0.6,-1.8+0.05) -- (-1.89+0.6,-1.8-0.05) node [below ]{};
	%\draw[gray] (-1.89+0.8,-1.8+0.05) -- (-1.89+0.8,-1.8-0.05) node [below ]{};
	%\draw[gray] (-0.89,-1.8+0.05) -- (-0.89,-1.8-0.05) node [below]{};
	%\useasboundingbox[black, font=\tiny] (-1.8,-1.8-0.3) -- (-0.6,-1.8-0.3) %node[pos=0.14,rotate=-30]{$R_C$} node[pos=0.45,rotate=-30]{$t_{AlSb}$} %node[pos=0.75,rotate=-30]{$t_{InAs}$}  node[pos=0.95,rotate=-30]{$t_{GaSb}$};
	%\node[options](node name)
	\end{tikzpicture}
	%\begin{tikzpicture}
	%\draw[black,thick]  -- (0,) node[draw]  {Placeholder};
	%\end{tikzpicture}
	\caption{(a) Evolution of the finite system eigenvalues with increasing axial disorder obtained from the BHZ/TB model. The color code is the same as in Fig.~\ref{fig_1Dbands_fin}(a). The states stemming from the splitting of the zero angular momentum end state are marked with the green and magenta dots. (b)  Probability density for  the marked states along the NW growth axis. (c) Evolution of the finite system eigenvalues with increasing angular disorder. Apart from disorder, the parameters are the same as in Fig.~\ref{fig_1Dbands_fin}(a).}
	\label{disorder}
\end{figure}
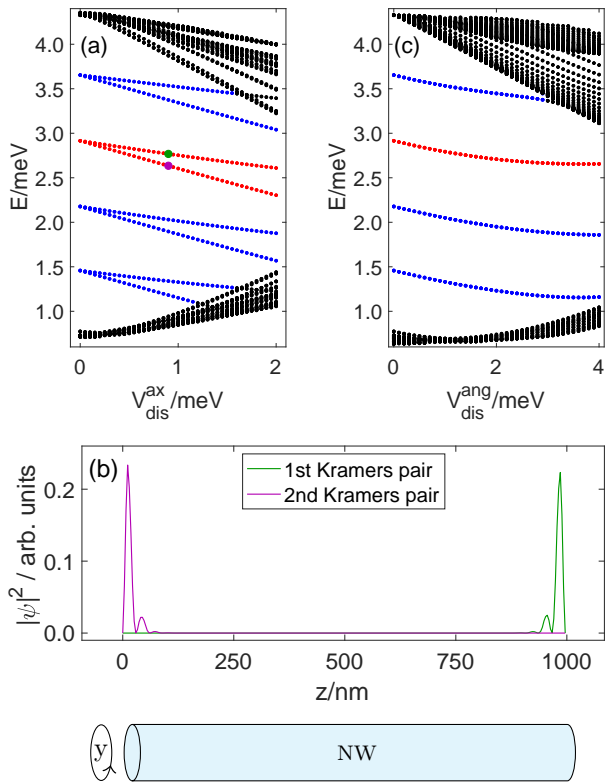

The robustness of the end states can be checked by including disorder effects in the NW BHZ/TB model. First we consider disorder in the axial (growth) direction ($V_{dis}^{ax}$) and no disorder in the angular direction. For each set of sites with the same axial coordinate $z$, identical disorder terms $H_{dis}^{ax}(z) = a(z) V_{dis}^{ax}\mathds{1}_{4 \times 4}$ are added to the corresponding $4 \times 4$ onsite submatrices  of the discretized version of Eq.~(\ref{TBhamtotal}), where each $a(z)$ is a random number from a uniform distribution in the interval $[-1,1]$. For $n$ sites in the $z$ direction, $n$ random $a(z) \equiv a(nd) = a_n$ are chosen ($d=5$~nm is the lattice constant). The same set of $a_n$'s is used for the different disorder strengths $V_{dis}^{ax}$. In Fig.~\ref{disorder}(a) we show the evolution of the eigenvalues of Fig.~\ref{fig_1Dbands_fin}(a) with increasing disorder strength, for a typical set of $a_n$'s. The color code is the same as in Fig.~\ref{fig_1Dbands_fin}(a). The most striking effect is the splitting of the end states' energies with increasing disorder. Since the disorder is time-reversal symmetry preserving, each eigenvalue remains Kramers degenerate also for $V_{dis}^{ax} \neq 0$. The Kramers-degenerate eigenvalues stemming from the splitting of the zero angular momentum end state energy for a disorder strength of $V_{dis}^{ax} = 0.9$~meV are marked with a green and a magenta dot. The wave functions' amplitude squared along the NW axis corresponding to the marked states are plotted in Fig.~\ref{disorder}(b).  The states remain localized at the ends of the wire for $V_{dis}^{ax} \neq 0$ and their energy splitting cannot be attributed to wave functions overlapping due to the finite wire length. The splitting is a signature of the lack of topological protection of the end states. %We conclude that the end states are not topologically protected since they are not robust against disorder.
% The disorder on each site is not affected by disorder on neighboring sites and the correlation length is equal to the TB lattice constant (for the finite NW $a = 5$~nm).
% and $\mathds{1}_{4 \times 4}$ is the $4\times4$ identity matrix
%The pseudo-randomness of $V(z)$ refers to the fact that for a given disorder strength the $V(z)$ terms added to the TB Hamiltonian are always the same.

Despite the above general conclusion, it turns out that some aspects of topological protection remain for the end states of the NW. This becomes evident if one considers a different type of disorder. Figure~\ref{disorder}(c) also shows the evolution of the eigenvalues of Fig.~\ref{fig_1Dbands_fin}(b) with increasing disorder strength, but this time the disorder is in the angular direction ($V_{dis}^{ang}$) and $V_{dis}^{ax} = 0$. The TB Hamiltonian is obtained with a procedure similar to the axial disorder case, but here the disorder terms $H_{dis}^{ang}(y) = b(y) V_{dis}^{ang}\mathds{1}_{4 \times 4}$ are added to sites with the same angular coordinate $y$. The end states do not split in this case and they remain 4-fold degenerate, undisturbed by disorder. 

%At large disorder the upper end state can be seen to be engulfed by the bulk states. The end states are thus robust against angular disorder since they do not split at a disorder strength large enough to close the gap.

%%%%%%%%%%%%%%%%%%%%%%%%%%%%%%%%%%%%%%%%%%%%%%
%%%%%%%%%%%%%%%%%%%%%%%%%%%%%%%%%%%%%%%%%%%%%%
%%%%%%%%%%%%%%%%%  Figure6	 %%%%%%%%%%%%%%%%%%%%%%%
%%%%%%%%%%%%%%%%%%%%%%%%%%%%%%%%%%%%%%%%%%%%%%
%%%%%%%%%%%%%%%%%%%%%%%%%%%%%%%%%%%%%%%%%%%%%%
\section{Conclusions}
We have studied InAs/GaSb core-shell-shell NWs by calculating their electronic band structure and wave functions. We find that, as in the case for QWs of the same materials, a hybridization gap can exist, and that the system hosts in-gap end states in this inverted regime. The end states are two-fold degenerate within each Kramers sector, and gap out when subject to axial disorder. However, disorder in the radial direction leaves the end states unaffected, as long as the bulk gap persists.

%%%%%%%%%%%%%%%%%%%%%%%%%%%%%%%%%%%%%%%%%%%%%%
%%%%%%%%%%%%%%%%%%%%%%%%%%%%%%%%%%%%%%%%%%%%%%
%%%%%%%%%%%%%%%  Acknowledgments - references %%%%%%%%%%%
%%%%%%%%%%%%%%%%%%%%%%%%%%%%%%%%%%%%%%%%%%%%%%
%%%%%%%%%%%%%%%%%%%%%%%%%%%%%%%%%%%%%%%%%%%%%%
\section{Acknowledgments}
This work was supported by NanoLund, by the Swedish Research Council (VR), and by the Knut and Alice Wallenberg Foundation (KAW).
Computational resources were provided by the Swedish National Infrastructure for Computing (SNIC) through Lunarc, the Center for Scientific and Technical Computing at Lund University.
%\appendix*
%\section{Details of the \kp calculations}

\bibliography{lib_paperII}

\end{document}